\documentclass[11pt]{article}
\usepackage{amsfonts,latexsym}

\newfont{\blackb}{msbm10 scaled\magstep1}

\newfont{\calig}{cmsy10 scaled\magstep1}

\def\text#1{\hbox{#1}}

\newtheorem{theorem}{Theorem}[section]

\newtheorem{remark}{Remark}[section]
\newtheorem{corollary}{Corollary}[section]

\def\nn{\nonumber}
\def\non{\nonumber\\}
\def\be{\begin{equation}}
\def\ee{\end{equation}}
\def\ben{\begin{displaymath}}
\def\een{\end{displaymath}}
\def\baa{\begin{eqnarray}}
\def\eaa{\end{eqnarray}}

\def\ba{\begin{array}}
\def\ea{\end{array}}

\def\3{\ss}

\def\d{\delta}
\def\e{\varepsilon}

\def\ka{\kappa}
\def\l{\lambda}

\def\s{\sigma}
\def\t{\tau}
\def\th{\vartheta}
\def\Th{\Theta}

\def\O{\Omega}

\def\phi{\varphi}

\def\eb{{\bf e}}

\def\Lh{\hat{\L}}

\def\Zf{{\cal Z}}

\def\B{{\bf B}}
\def\C{\mathbb{C}}
\def\Z{\mathbb{Z}}

\def\t0{\Theta_0}

\def\m{{\bf m}}
\def\la{\label}
\def\Ref{\ref}
\def\c{\cite}
\def\f{\frac}
\def\L{{\cal L}}
\def\p{\partial}
\def\pb{{\bf p}}
\def\qb{{\bf q}}

\def\Ref#1{(\ref{#1})}
\def\tr{{\rm tr}}
\def\0{S}
\def\1{T}

\def\log{\ln}

\def\Zf{{\cal Z}}
\def\Lh{\hat{\L}}

\def\B{{\bf B}}
\def\C{\mathbb{C}}
\def\Z{\mathbb{Z}}

\def\t0{\Theta_0}

\def\m{{\bf m}}
\def\la{\label}
\def\Ref{\ref}
\def\c{\cite}
\def\f{\frac}
\def\L{{\cal L}}
\def\p{\partial}

\def\Ref#1{(\ref{#1})}
\def\tr{{\rm tr}}
\def\0{S}
\def\1{T}

\def\log{\ln}

\def\res{{\rm res}}
\def\hPsi{{\widehat{\Psi}}}
\def\htau{{\widehat{\tau}}}
\def\tht{{\widehat{t}}}
\def\hA{{\widehat{A}}}
\def\hG{{\widehat{G}}}
\def\hH{{\widehat{H}}}
\def\hM{{\widehat{M}}}
\def\i{{\rm i}}
\def\sl{{\mathfrak{sl}}}

\def\Lh{{\widehat{{\cal L}}}}
\def\ft#1#2{{\textstyle {\frac{#1}{#2}} }}
\def\cO{{\cal{O}}}

\def\A{{\alpha}}
\def\Ab{{\beta}}
\def\Ac{{\gamma}}

\newcommand{\qed}{\hfill$\Box$}

\begin{document}

\begin{center}{\LARGE Isomonodromic deformations in genus zero and one:
algebrogeometric solutions  and Schlesinger transformations}\\
\vskip1.0cm

{\large D.~A.~Korotkin}\footnote{E-mail: korotkin@aei-potsdam.mpg.de}\\
\vskip0.5cm
Max-Planck-Institut f\"ur Gravitationsphysik, \\
Am M\"uhlenberg  1, D-14476 Golm, Germany

\end{center}

\section{Introduction}

Here we review some recent developments in the theory of isomonodromic deformations on Riemann sphere 
and elliptic curve. For both cases we show how to derive 
 Schlesinger transformations together with their action 
on tau-function, and construct classes of solutions in terms of multi-dimensional 
theta-functions. 

The theory of isomonodromic deformations of ordinary matrix
differential equations of the type 
\be\la{A} \f{d\Psi}{d\l}= A(\l)\,\Psi
\;, 
\ee 
where $A(\l)$ is a matrix-valued meromorphic function on
$\overline{\C}$, is a classical area intimately related to the matrix
Riemann-Hilbert problem on the Riemann sphere.  Over the last 20 years
this has become a powerful tool in areas like soliton theory,
statistical mechanics, theory of random matrices, quantum field theory
etc. The main object associated with the isomonodromic deformation
equations is the so-called $\tau$-function.

After the
classical work of Schlesinger \cite{Schl12} the important contributions to
the development of the subject were made in the papers of Jimbo, Miwa
and their collaborators in the early 80's
\cite{JiMiMoSa80,JiMiUe81,JimMiw81b,JimMiw81c}.

There are only a few cases where the matrix Riemann-Hilbert problem
may be solved explicitly in terms of known special functions.

However, as was
already discovered by Schlesinger himself, there exists a large class
of transformations which allow to get an infinite chain of new
solutions starting from the known ones. They share the characteristic
feature that they shift the eigenvalues of the residues of the
connection $A(\l)$ in \Ref{A} by integer or half-integer values, thus
changing the associated monodromies by sign only. These
transformations -- nowadays called Schlesinger transformations -- were
systematically studied in \cite{JimMiw81b,JimMiw81c}. In particular,
it turns out that being written in terms of the $\tau$-functions the
superposition laws of these transformations provide a big supply of
discrete integrable systems.

Recently in papers \cite{KitKor98,DIKZ99} it was solved a class of $2\times 2$ 
Riemann-Hilbert problems with arbitrary off-diagonal monodromy matrices in terms of 
multidimensional theta-functions. The equations for
$\tau$-function were integrated in the paper  \cite{KitKor98} to give the following result:
$$
\tau(\{\l_j\})=[\det{\cal A}]^{-\f 12}
\prod\limits_{j<k}(\l_j-\l_k)^{-\frac 18}\Theta\left[^\pb_\qb\right](0|\B)\;.
$$
where all the objects associated to auxiliary hyperelliptic curve are defined below in Sect.2.3.

The natural question of generalizing the theory of isomonodromic
deformations on the sphere to higher genus surfaces was addressed by
several authors. Here, we mention the contributions of Okamoto
\cite{Okam71,Okam79} and Iwasaki \cite{Iwas91}.

For the case of the torus, recently two different explicit forms of equations of
isomonodromic deformations were proposed. In work of the author and Samtleben \cite{KorSam97}
it were studied isomonodromic deformations of
non-singlevalued meromorphic connection on the torus whose ``twists''
(which determine the transformation of the connection $A(\l)$ with
respect to tracing along basic cycles of the torus) vary with respect
to the deformation parameters. The isomonodromic deformation equations
for these connections hence contain transcendental dependence on the
dynamical variables, which makes it difficult to analyse this system
in a way analogous to the Schlesinger system on the sphere.  On the
other hand, Takasaki \cite{Taka98} considered connections on the torus
whose twists remain invariant with respect to the parameters of
deformation. In Takasaki's form, the equations of isomonodromic
deformations have already the same degree of non-linearity as the
ordinary Schlesinger system.

In the paper \cite{KMS99} it were constructed transformations of  Schlesinger 
type for elliptic isomonodromic deformations in Takasaki form, and it was derived the action 
of these transformation on elliptic version of $\tau$-function.
Here we review these results, and, in addition, present the generalization of results of the
paper \cite{KitKor98} to elliptic case. We show how to solve certain class of 
Riemann-Hilbert problems on the torus in terms of Prym theta-functions. In turn, this allows
to construct a class of algebro-geometric solutions of elliptic Schlesinger system.

In sect.2 we introduce Schlesinger system on the Riemann sphere. For $2\times 2$ case we 
discuss elementary Schlesinger transformations together with their action on $\tau$-function
and, following \cite{KitKor98},
 derive class of algebro-geometric solutions of the Schlesinger system in terms of theta-functions of
auxiliary hyperelliptic curve. In sect.3 we describe equations of elliptic isomonodromic
deformations with constant twists \cite{Taka98}, and, following \cite{KMS99}, construct elliptic
version of elementary Schlesinger transformations. The new result of this paper 
- the construction of
algebro-geometric solutions of elliptic Schlesinger system in terms of Prym theta-functions -
is presented in sect.3.

\section{Schlesinger system on the Riemann sphere: Schlesinger transformations and 
algebro-geometric solutions}

\subsection{Schlesinger system}

Consider the following ordinary linear differential equation (\ref{A}) for a
matrix-valued function $\Psi(\lambda)\in SL(2,\C)$ and 
\be
A(\l)= 
\sum_{j=1}^N\f{A_j}{\l-\l_j},
\la{ls}\ee
where the residues $A_j\in \sl(2,\C)$ are independent of
$\l$. Regularity at $\l=\infty$ requires
\be
\sum_{j=1}^N A_j ~=~ 0\;,
\ee
and allows to further impose the initial condition
$\Psi(\l\!=\!\infty)=I$. 
The matrix $\Psi(\l)$ defined in this way lives on the
universal covering $X$ of $\C P^1\setminus\{\l_1,\dots,\l_N\}$. Its
asymptotical expansion near the singularities $\l_j$ is given by
\be
\Psi(\l)= G_j\Psi_j\cdot \,(\l-\l_j)^{T_j}\, C_j\;,
\la{asymp}\ee
with $G_j\,,C_j\,\in SL(2,\C)$ constant,
$\Psi_j=I+\cO(\l\!-\!\l_j)\in SL(2,\C)$ holomorphic around
$\l\!=\!\l_j$, and where $T_j$ is a traceless diagonal matrix with
eigenvalues $\pm t_j$. The residues $A_j$ of \Ref{ls} are encoded in
the local expansion as 
\be
A_j =G^{\phantom{1}}_j \,T^{\phantom{1}}_j\, G_j^{-1}\;.
\la{Aj}
\ee

Upon analytical continuation around $\l\!=\!\l_j$, the function
$\Psi(\lambda)$ in $\C P^1\setminus\{\l_1,\dots,\l_N\}$ changes by
right multiplication with some monodromy matrices $M_j$
\baa
\Psi(\l) &\to&  \Psi(\l)\,M_j\;,\la{Mj}\\[4pt]
M_j&=&C_j^{-1}\, e^{2\pi i T_j}\, C^{\phantom{1}}_j\;.\nn
\eaa
In the sequel we shall consider the generic case when none of $t_j$ is
integer or half-integer.
 
The assumption of independence of all monodromy matrices $M_i$ of the
positions of the singularities $\l_j$: $ \p M_i/\p\l_j=0$ is called
the isomonodromy condition; it implies the following dependence of
$\Psi(\l)$ on $\l_j$
\be
\f{\p\Psi}{\p\l_j}=-\f{A_j}{\l-\l_j}\;\Psi\;,
\la{ls1}\ee
as follows from \Ref{asymp} and normalization of $\Psi(\l)$ at
$\infty$. Compatibility of (\ref{A}) and (\ref{ls1}) then is
equivalent to the classical Schlesinger system \c{Schl12}:  
\be
\frac{\partial A_j}{\partial\l_i}=
\frac{[A_j,A_i]}{\l_j-\l_i}\;,\quad i\neq j\;,\qquad
\frac{\partial A_j}{\partial\l_j}=
-\sum_{i\neq j}
\frac{[A_j,A_i]}{\l_j-\l_i}\;,
\label{sch}
\ee
describing the dependence of the residues $A_j$ on the
$\l_i$. Obviously, the eigenvalues $t_j$ of the $A_j$ are integrals of
motion of the Schlesinger system. The functions $G_j$  have
the following dependence on $\l_j$: \cite{JiMiUe81}:
\be
\f{\p G_j}{\p \l_i} = \f{A_i G_j}{\l_i-\l_j}\;,\quad i\neq j\;,\qquad  
\f{\p G_j}{\p \l_j} = -\sum_{i\neq j}\f{A_i G_j}{\l_i-\l_j} \;,
\la{Geq}\ee
which obviously implies (\ref{sch}).

To introduce the notion of the $\tau$-function for the Schlesinger
system, one notes that \Ref{sch} is a multi-time Hamiltonian system
\c{JiMiMoSa80} with respect to the Poisson structure on the residues
$A_j$
\be
\left\{A_i^\A\,, A_j^\Ab\;\right\}= \d_{ij}\,\e^{\A\Ab\Ac}\,A_j^\Ac\;,
\la{Poisson}\ee
($\A,\Ab,\Ac$ denoting $\sl(2)$ algebra indices with the completely
antisymmetric structure constants $\e^{\A\Ab\Ac}$) and Hamiltonians
\be\la{Hs}
H_{i}~=~ \f{1}{4\pi\i}\oint_{\l_i} \tr A^2(\l)\,d\l ~=~
\ft12\,\sum_{j\neq i}\f{\tr A_i A_j}{\l_j\!-\!\l_i} \;.
\ee
Explicitly, \Ref{sch} takes the form
\be
\f{\p A_j}{\p \l_i}= \{H_i,  A_j\}\;,
\ee
and all the Hamiltonians $H_j$ Poisson-commute.

The $\tau$-function $\tau (\{\l_j\})$ of the Schlesinger system then
is defined as the generating functions of the Hamiltonians
\be
\f{\p\ln\tau}{\p\l_j} ~=~ H_j \;,
\la{tauHj}
\ee
where compatibility of these equations follows from \Ref{sch}. This
$\tau$-function is closely related to  the Fredholm determinant of a certain integral
operator associated to the Riemann-Hilbert problem (see
\cite{HarIts97} for details).

\subsection{Schlesinger transformations on the Riemann sphere}
Schlesinger transformations are symmetry transformations of the
Schlesinger system (\ref{sch}) which map a given solution
$\left\{A_j(\{\l_i\})\right\}$ to another solution
$\left\{\hA_j(\{\l_i\})\right\}$ with the same number and positions of
poles $\l_j$ such that the related eigenvalues $t_j$ are shifted by
integer or half-integer values $t_j\to t_j\!+\!n_j/2\;,\;\; n_j\in
\Z$. The monodromy matrices $M_j$ hence remain invariant or change
sign under this transformation. 
We shall restrict ourselves to elementary Schlesinger transformations,
which change only two $t_j$'s, say, $t_k$ and $t_l$ for $k\neq l$ by
$\pm 1/2$. The transformed variables will be denoted by
$\hPsi,\,\hA_j,\, \tht_j$, etc. Without loss of generality we consider
the case
\be\la{tj}
\tht_j~=~ \left\{ \begin{array}{ll} t_j\!+\!\ft12\;\quad &
                               \mbox{for~} j=k,l \\  
                               t_j &\mbox{else} 
                               \end{array} \right. \;.
\ee
Our presentation here mainly follows \cite{Kita97}. For the transformed
function $\hPsi$ we make the ansatz
\be
\hPsi(\l) ~=~ F(\l)\,\Psi(\l) \;,
\la{Psih}\ee
with 
\be
F(\l)~=~ \sqrt{\f{\l-\l_k}{\l-\l_l}}\; S_+  +  
         \sqrt{\f{\l-\l_l}{\l-\l_k}}\; S_-  \;,
\la{Fsphere}
\ee
where the matrices $S_\pm$ do not depend on $\l$ and are uniquely
determined by \cite{Kita97}:
\be
S_\pm^2=S_\pm\;,\qquad S_++S_-=I\;,\qquad
S_+\,G_l^1= S_-\,G_k^1 =0\;.
\la{eqS}\ee
By $G_j^{\alpha}$ here we denote the $\alpha$-th column of the matrix
$G_j$ ($\alpha=1,2$). Combining the columns $G_k^1$ and $G_l^1$ into a
$2\times 2$ matrix 
\be
G=\left(G_k^1\,,  G_l^1\right), 
\la{Gkl}\ee
we can deduce from (\ref{eqS}) the following simple 
formula for $S_\pm$:
\be
S_\pm= G\, P_\pm\,  G^{-1} \;,
\la{SGkl}\ee
with projection matrices
\ben
P_+ =\left(\ba{cc} 1 & 0\\
                   0 & 0  \ea   \right)\;,\hskip1.3cm
P_- =\left(\ba{cc} 0 & 0\\
                   0 & 1  \ea   \right) \;.
\la{ppm}\een
It is easy to check using the local expansion of $\Psi$ at the
singularities $\l_j$ (\ref{asymp}) and the defining relations for
$S_\pm$ (\ref{SGkl}) that the transformed function $\hPsi$ at $\l_j$
has a local expansion of the form (\ref{asymp}) with the same matrices
$C_j$ and the desired transformation \Ref{tj} of the $t_j$. The
matrices $G_j$ change to new matrices $\hG_j$.  Thus, $\hPsi$
satisfies the system  
\be \la{lsh}
\f{\p\hPsi}{\p\l}= \sum_{j=1}^N\f{\hA_j}{\l\!-\!\l_j}\;\hPsi\;,
\qquad
\f{\p\hPsi}{\p\l_j}=-\f{\hA_j}{\l\!-\!\l_j}\;\hPsi\;,
\ee
where the functions $\hA_j(\{\l_i\})$ build a new solution of the
Schlesinger system (\ref{sch}).

On the level of the residues $A_j$, the form of the Schlesinger
transformation is not very transparent; however, it turns out that the
associated $\tau$-function transforms in a rather simple way.
Namely, for $\hPsi$ we find
\be
\tr\,\hA^2 =  
\tr\,A^2 + 
2\,\tr\left[ F^{-1}\f{dF}{d \l}\; A\right]+
\tr\left[ F^{-1}\f{dF}{d\l}\right]^2 \;.
\la{trAh}\ee
For example, the Hamiltonians $H_j$ for $j\neq k,l$ transform as follows:
\ben
\hH_j-H_j =\left(\f{1}{\l_j\!-\!\l_k}-\f{1}{\l_j\!-\!\l_l}\right)\,
\tr\left[A_j S_+\right] 
=\f{\tr\left[A_j G P_+ G^{-1}\right]}{\l_j-\l_k} + 
   \f{\tr\left[A_j G P_- G^{-1}\right]}{\l_j-\l_l} 
\een
\ben
=\tr\left[\;\f{\p G}{\p\l_j}\;G^{-1}\;\right] 
\een
according to (\ref{Geq}).
Hence the transformed $\tau$-function $\htau$ is given by 
$\htau= f(\l_k,\l_l) \det G\,\cdot\, \tau$ with some function
$f(\l_k,\l_l)$ to be determined from the transformation of $H_k$,
$H_l$. Taking into account the transformation of
Hamiltonians $H_k$ and $H_l$ following from (\ref{trAh}) we find
 the following formula describing
the action of elementary Schlesinger transformation \Ref{tj} on the
$\tau$-function:  
\be
\htau\left(\{\l_j\}\right) ~=~ \left\{(\l_k\!-\!\l_l)^{-1/2}\; 
\det G \right\}\,\cdot\,\tau\left(\{\l_j\}\right) \;.
\la{tauh}\ee
\bigskip

Other elementary Schlesinger transformations like
 may be obtained in a similar way by building the matrix $G$
from $G^1_k$ and $G^2_l$ instead of \Ref{Gkl}, etc..  Moreover, all
such transformations with different $k$ and $l$ may be superposed to
get the general Schlesinger transformation which simultaneously shifts
an arbitrary number of the $t_j$ by some integer or half-integer
constants.  These general transformations were in detail studied in
\cite{JiMiUe81,JimMiw81b,JimMiw81c}.

\subsection{Algebro-geometric solutions of Schlesinger system}

Let us take $N=2g+2$ and  introduce the
hyperelliptic curve $\L$ of genus $g$ by the equation
\be
w^2=\prod_{j=1}^{2g+2}(\l-\l_j)
\la{L}\ee
with branch cuts $[\l_{2j+1},\l_{2j+2}]$. Let us choose the canonical basis of cycles 
$(a_j,b_j),\;j=1,\dots, g$ such that the cycle $a_j$ encircles the branch cut   $[\l_{2j+1},\l_{2j+2}]$.
Cycle $b_j$ starts from one bank of branch cut $[\l_1,\l_2]$, goes to the second sheet through
he branch cut  $[\l_{2j+1},\l_{2j+2}]$, and comes back to another bank of the  branch cut $[\l_1,\l_2]$.

The dual basis of  holomorphic 1-forms on $\L$
are given by
$\f{\l^{k-1}d\l}{w},\;\;k=1,\dots,g$.

Let us introduce two $g\times g$ matrices of $a$- and $b$-periods
of these 1-forms: 
\be
{\cal A}_{kj}=\oint_{a_j}\f{\l^{k-1}d\l}{w},\;\;\;\;\;\;\;
{\cal B}_{kj}=\oint_{b_j}\f{\l^{k-1}d\l}{w}.
\la{AB}\ee
The holomorphic 1-forms
\be
dU_k=\f{1}{w}\sum_{j=1}^g ({\cal A}^{-1})_{kj} \l^{j-1} d \l
\la{dUk}\ee
satisfy the normalization conditions
$\oint_{a_j} dU_k=\delta_{jk}$.

The matrices ${\cal A}$ and ${\cal B}$ define the symmetric
$g\times g$ matrix of $b$-periods of the curve $\L$:
$\B= {\cal A}^{-1}{\cal B}$.

Let us cut the curve $\L$ along all basic cycles to get the fundamental polygon   $\Lh$.
For any meromorphic 1-form $dW$ on $\L$ we can define the integral $\int_Q^P dW$,
where the integration contour lies inside of $\Lh$ (if $dW$ is meromorphic,
the value of this integral might also depend on the choice of integration
contour inside of $\Lh$).
The vector of Riemann constants corresponding to our choice of the
initial point of this map is given by the formula (see \cite{Fay}) $K_j=\f{j}{2}+\f{1}{2}\sum_{k=1}^g \B_{jk}$.

The characteristic with components $\pb\in\C^g/2\C^g$,
$\qb\in\C^g/2\C^g$ is called half-integer characteristic: the half-integer
 characteristics are in one-to-one correspondence with the half-periods
$\B\pb+\qb$. To any half-integer characteristic we can assign parity which by definition coincides
 with the parity  of the scalar product $4\langle\pb,\qb\rangle$.

The odd characteristics which will be of importance for us in the sequel correspond to any given subset $S=\{\l_{i_1},\dots,\l_{i_{g-1}}\}$ of $g-1$ arbitrary non-coinciding branch points. The odd half-period associated to the subset $S$ is given by
\be
\B\pb^S+\qb^S= \sum_{j=1}^{g-1}\int_{\l_{1}}^{\l_{i_j}} d U -K
\la{odd}
\ee
where $dU=(dU_1,\dots,dU_g)^t$.
Denote by
 $\Omega\subset\C$ the neighbourhood of the infinite point $\l=\infty$,
such that $\Omega$ does not overlap with projections of all basic cycles
on $\l$-plane.
Let the $2\times 2$ matrix-valued function $\Phi(\l)$ be defined in the
domain $\O$ of the first sheet of $\L$  by the following formula,
\be
\Phi(\l\in\O_\l)=\left(\ba{cc}\phi(\l)\;\;\;\;\; \phi(\l^*)\\
                  \psi(\l)\;\;\;\;\; \psi(\l^*)\ea\right),
\la{Phi}\ee
where functions $\phi$ and  $\psi$ are defined in the fundamental
polygon $\Lh$ by the formulas:
\be
\phi(\l)=\Th\left[^\pb_\qb\right]\left(\int_{\l_1}^\l dU + 
\int_{\l_1}^{\l_\phi}dU\Big|\B\right)\Th\left[^{\pb^\0}_{\qb^\0}\right]\left(\int_{\l_\phi}^\l dU
\Big|\B\right),
\la{phi1}
\ee
\be
\psi(\l)=\Th\left[^\pb_\qb\right]\left(\int_{\l_1}^\l dU + 
\int_{\l_1}^{\l_\psi}dU\Big|\B\right)\Th\left[^{\pb^\0}_{\qb^\0}\right]\left(\int_{\l_\psi}^\l dU
\Big|\B\right),
\la{psi1}
\ee
with two arbitrary (possibly $\{\l_j\}$-dependent) points $\l_{\phi}$,
$\l_\psi\in\L$ and arbitrary constant complex characteristic $\left[^\pb_\qb\right]$; 
$*$ is the involution on $\L$ interchanging the sheets.
An odd theta characteristic $\left[^{\pb^\0}_{\qb^\0}\right]$ corresponds to an arbitrary subset $S$ of $g-1$ branch points via  {\rm Eq.~(\ref{odd})}.

Since domain $\O$ does not overlap with projections of all basic cycles
of $\L$ on $\l$-plane, domain $\O^*$ does not overlap with the boundary
of $\Lh$, and functions $\phi(\l^*)$ and $\psi(\l^*)$ in (\ref{Phi}) are
uniquely defined by (\ref{phi1}), (\ref{psi1}) for $\l\in\O$.

Now choose some sheet of the universal covering $X$,
define new function $\Psi(\l)$ in subset $\Omega$ of this sheet
by the formula
\be
\Psi(\l\in\Omega)=\sqrt{\f{\det \Phi(\infty^1)}{\det \Phi(\l)}}\Phi^{-1}(\infty^1)\Phi(\l)
\la{Psi}\ee
and extend on the rest of  $X$ by analytical continuation.

Function  $\Psi(\l)$ (\ref{Psi}) transforms as follows with respect
to the tracing around basic cycles of $\L$ (by $T_{a_j}$ and $T_{b_j}$
we denote corresponding  operators of analytical continuation):
\ben
T_{a_j}[\Psi(\l)]=\Psi(\l) e^{2\pi i p_j\sigma_3}\;; \hskip1.0cm
T_{b_j}[\Psi(\l)]=\Psi(\l)  e^{-2\pi i q_j\sigma_3}  \;
\een

The following statement proved in the paper \cite{KitKor98} claims that function $\Psi$ 
satisfies condition of isomonodromy, and, therefore, provides a class of solutions of Schlesinger system:

\begin{theorem}\la{theoPsi}
Let $\pb,\qb\in \C^g$ be an arbitrary set of $2g$ constants such that characteristic
$\left[^\pb_\qb\right]$ is not half-integer. Then:
\begin{enumerate}
\item
Function $\Psi(Q\in X)$ defined by (\ref{Psi}) is independent of $\l_\phi$
and $\l_\psi$, and satisfies the linear system (\ref{ls}) with
\be
A_j\equiv {\rm res}|_{\l=\l_j} \left\{\Psi_\l\Psi^{-1}\right\},
\ee
which in turn solve the Schlesinger system (\ref{sch}).
\item
Monodromies (\ref{Mj}) of $\Psi(\l)$ around points $\l_j$ are given by
\be
M_j= \left(\ba{cc} 0 & -m_j \\m_j^{-1} & 0 \ea\right)\;,
\la{Mj1}\ee
where constants $m_j$ may be  expressed in terms of $\pb$ and $\qb$
as follows:
\ben
m_1 = i\hskip1.0cm 
m_2 =  i\exp\{-2\pi i\sum_{k=1}^g p_k\}
\een
\ben
m_{2j+1}=  -i\exp\{2\pi i q_j -2\pi i \sum_{k=j}^g p_k\}
\een
\ben
m_{2j+2}=  i\exp\{2\pi i q_j - 2\pi i\sum_{k=j+1}^g p_k\}
\een
for $j=1,\dots, g$.
\item
The $\tau$-function, corresponding to solution  (\ref{Aj}) of the
Schlesinger system, has the following form:
\be
\tau(\{\l_j\})=[\det{\cal A}]^{-\f 12}
\prod\limits_{j<k}(\l_j-\l_k)^{-\frac 18}\Theta\left[^\pb_\qb\right](0|\B)\;.
\la{tau}\ee
\end{enumerate}
\end{theorem}

\section{Elliptic isomonodromic deformations: Schlesinger transformations and algebro-geometric solutions}

\subsection{Isomonodromic deformations on the torus}

Consider the elliptic curve $E$ with periods $1$ and $\mu$ together
with the canonical basis of cycles $(a,b)$. A (naive) straightforward
generalization of the idea of isomonodromic deformations from the
complex plane to the torus $E$ runs into difficulties related to the
absence of meromorphic functions on the torus with just one simple
pole. An independent variation of the simple poles of a meromorphic
connection $A$ on the torus preserving the monodromies around the
singularities and basic cycles is impossible for the following simple
reason. Existence of such a deformation would imply a version of
(\ref{ls1}) with the function $\f{A_j}{\l-\l_j}$ on the r.h.s.~being
substituted by a meromorphic function with only one simple pole on the
torus, which gives rise to the contradiction.  Therefore, one of the
underlying assumptions has to be relaxed.

E.g.~one may consider the case where not all the poles of the
connection $A$ are varied independently. Another possibility is the
assumption that some of the poles of $A$ are of order higher then one
\cite{Okam79}. A third alternative which we shall consider here, is to
relax the condition of single-valuedness of the connection $A$ on $E$
and assume that $A$ has ``twists'' with respect to analytical
continuation along the basic cycles $a$ and $b$, i.e.
\ben
A(\l+1) = Q A(\l) Q^{-1}\;,\qquad
A(\l+\mu) = R A(\l) R^{-1} \;,
\een
where the matrices $Q,\,R$ do not depend on $\l$. By a gauge
transformation of the form $A\to S A S^{-1} + dS S^{-1}$ with $S$
holomorphic but possibly multi-valued, one may bring the connection
into a form where $Q=I$ and $R=e^{\ka \sigma_3}$,  where $\sigma_\A$
denote the Pauli matrices: 
\ben
\sigma_1 =\left(\ba{cc} 0 & 1\\
                       1 & 0  \ea   \right)\;,\qquad
\sigma_2 =\left(\ba{cc} 0  & \i\\
                 -\i & 0       \ea\right)\;,\qquad
\sigma_3 =\left(\ba{cc} 1  & 0\\
                       0 & -1      \ea\right)\;.
\een
The equations of isomonodromic deformations with this choice of the
twist were considered in \cite{KorSam97} where the multi-valuedness of
$A$ had a natural origin in the holomorphic gauge fixing of
Chern-Simons theory on the punctured torus. The resulting equations
however are rather complicated in comparison with the Schlesinger
system on the sphere.  This is due to the fact that the twist $\ka$
itself becomes a dynamical variable -- i.e.~changes under
isomonodromic deformations -- and in generic situation has a highly
non-trivial $\l_j$-dependence.  Therefore, instead of being bilinear
with respect to the dynamical variables, this Schlesinger system on
the torus becomes highly transcendental.
  
An alternative form of the elliptic Schlesinger system was proposed by
Takasaki \cite{Taka98} who considered the restriction $Q=\sigma_3$,
$R=\sigma_1$, related to the classical limit of Etingof's elliptic
version of the Knizhnik-Zamolodchikov-Bernard system on the torus
\c{Etin94}. This choice of fixing the twists turns
out to be compatible with the isomonodromic deformations equations,
therefore essentially simplifying the dynamics as compared to
\cite{KorSam97}.  It results into studying isomonodromic deformations
of the system
\baa
\f{d\Psi}{d\l}&=& A(\l)\, \Psi\;, \la{lstor} \\
A(\l)&\equiv&\sum_{j=1}^N \sum_{\A=1}^3 A^{\A}_j \,w_\A(\l\!-\!\l_j) \sigma_\A
\;, \nn
\eaa
with $\l\in \C$. Functions $w_\A$ on the torus are defined in Appendix 
(see \Ref{wj}). The connection
$A(\l)$ obviously has only simple poles on $E$ and the following twist
properties, cf.~\Ref{perw}  
\be
A(\l+1)= \sigma_3 \,A(\l)\, \sigma_3 \;,\qquad
A(\l+\mu)= \sigma_1\, A(\l) \,\sigma_1\;.
\la{Taktwist}\ee
Since the residues of all $w_\A$ at $\l=0$ coincide, the residue of
$A(\l)$ at $\l_j$ is
\ben
A_j \equiv \sum_{\A} A^{\A}_j \, \sigma_\A\;.
\een                                                                           
As in the case of the Riemann sphere, the function $\Psi$ has regular
singularities at $\l=\l_j$ with the same local properties
\Ref{asymp}--\Ref{Mj}. The twist properties of $\Psi$ take the
form 
\be
\Psi(\l+1) = \sigma_3  \Psi(\l) M_a \hskip1.0cm
\Psi(\l+\mu) = \sigma_1  \Psi(\l) M_b \;,
\la{perPsi}\ee
with monodromy matrices $M_a$, $M_b$ along the basic cycles of the
torus.  Moreover, as in the case of Riemann sphere, $\Psi(\l)$  
has monodromies $M_j$ around the singularities $\l_j$.

The isomonodromy condition on the torus requires that all monodromies
$M_j$, $M_a$ and $M_b$ are independent of the positions of
singularities $\l_j$ and the module $\mu$ of the torus. As on the
Riemann sphere this implies that the function
${\p\Psi}/{\p\l_j}\,\Psi^{-1}$ has the only simple pole at
$\l\!=\!\l_j$ with residue $-A_j$. In addition, it has the following
twist properties
\baa
\f{\p\Psi}{\p{\l_j}}\,\Psi^{-1} (\l+1)&=& 
\sigma_3\,\f{\p\Psi}{\p\l_j}\,\Psi^{-1} (\l)\,\sigma_3 \;,
\non
\f{\p\Psi}{\p\l_j}\,\Psi^{-1} (\l+\mu)&=& 
\sigma_1\,\f{\p\Psi}{\p\l_j}\,\Psi^{-1} (\l)\,\sigma_1 \;.
\nn
\eaa
Therefore,
\be
\f{\p\Psi}{\p\l_j} = - \sum_{\A=1}^3 A^{\A}_j\, w_\A(\l-\l_j) \sigma_\A\, 
\Psi\;.
\la{Psigjtor}\ee
To derive the equation with respect to module $\mu$ we observe that
$\p\Psi/\p\mu\,\Psi^{-1}$ is holomorphic at $\l\!=\!\l_j$ (but not at
$\l=\l_j\!+\!\mu$\,) and has twist properties  
\baa
\f{\p\Psi}{\p\mu}\,\Psi^{-1} (\l+1) &=&
\sigma_3\, \f{\p\Psi}{\p\mu}\,\Psi^{-1} (\l)\,\sigma_3 \;,
\non
\f{\p\Psi}{\p\mu}\,\Psi^{-1} (\l+\mu) &=&
\sigma_1\,\left(\f{\p\Psi}{\p\mu}\,\Psi^{-1} (\l)- 
\f{\p\Psi}{\p\l}\,\Psi^{-1} (\l)\right)\,\sigma_1 \;.
\nn
\eaa
Taking into account the periodicity properties of the functions
$\Zf_\A$ (\ref{perZ}), this hence implies
\be
\f{\p\Psi}{\p\mu}= \sum_{j=1}^N \sum_{\A=1}^3 A^{\A}_j\, 
\Zf_\A (\l-\l_j)\sigma_\A\;\Psi \;.
\la{Psitau}\ee
The compatibility conditions of the equations (\ref{lstor}),
(\ref{Psigjtor}) and  (\ref{Psitau}) then yield the $\l_i$ and $\mu$
dependence of the residues $A_j$. The result is summarized in the
following 
\begin{theorem}
{\rm \c{Taka98}} Isomonodromic deformations of the system
(\ref{lstor}) are described by the following elliptic version of the
Schlesinger system: 
\baa
\f{d A_j}{d \l_i}&=& \left[\,A_j\,, \sum_{\A=1}^3 A_i^\A\, 
w_\A(\l_j\!-\!\l_i)\,\sigma_\A\,\right]\;,\qquad i\neq j \;,\la{Se1}\\[2pt]
\f{d A_j}{d \l_j}&=&-\sum_{i\not=j}
\left[\,A_j\,, \sum_{\A=1}^3 A_i^\A\, w_\A (\l_j\!-\!\l_i)\,
\sigma_\A\,\right] \;,\non[2pt]
\f{d A_j}{d \mu}&=& 
-\sum_{i=1}^N \left[\,A_j\,, \sum_{\A=1}^3 A_i^\A\, 
\Zf_\A (\l_j\!-\!\l_i)\,\sigma_\A\,\right] \;.
\nn
\eaa
\end{theorem} 
\qed

\medskip

The corresponding equations for the matrices $G_j$ from (\ref{asymp})
take a form analogous to the equations (\ref{Geq}) on the Riemann
sphere: 
\be
\f{\p G_j}{\p\l_i}=\sum_\A A_i^\A\, w_\A(\l_i\!-\!\l_j)\,\s_\A\; G_j\;,\qquad
\f{\p G_j}{\p\l_j}= -\sum_{i=1}^N \sum_\A  
A_i^\A\, w_\A(\l_i\!-\!\l_j)\,\s_\A\; G_j \;.
\la{Gell}\ee
The system (\ref{Se1}) admits a multi-time Hamiltonian formulation
with respect to the Poisson structure \Ref{Poisson} on the residues
\be
\{A_i^\A, A_j^\Ab\}= \d_{ij}\,\e^{\A\Ab\Ac}\,A_j^\Ac \;.
\la{rmatr}\ee

The Hamiltonians describing deformation with respect to the variables
$\l_i$ and to the module $\mu$ of the torus are respectively given by
\baa
H_i &=& \f{1}{4\pi i}\oint_{\l_i}\tr A^2(\l)d\l ~=~
\sum_{j\neq i} \sum_{\A} A_j^\A A_i^\A \, w_\A(\l_j-\l_i) \;,
\la{Hjcon}\\[6pt]
H_\mu &=& -\f{1}{2\pi i}\oint_a \tr A^2(\l)d\l ~=~
- \sum_{i,j} \sum_{\A}   A_i^\A A_j^\A \, {\cal{Z}}_\A(\l_i-\l_j) 
\;. \la{Hmucon}
\eaa

The representation of
$H_\mu$ as contour integral along the basic $a$-cycle in
(\ref{Hmucon}) was derived in \cite{KMS99}. All
Hamiltonians Poisson-commute as a direct consequence of \Ref{rmatr}. 

\medskip

The $\tau$-function of the elliptic
Schlesinger system \Ref{Se1} is defined as generating function
$\tau\left(\{\l_j\},\mu\right)$ of the Hamiltonians   
\be
\f{\p\ln\tau}{\p\l_j} = H_j\;,\qquad
\f{\p\ln\tau}{\p\mu} = H_\mu \;;
\la{tauell}\ee 
it is uniquely determined up to an arbitrary
$(\mu,\{\l_j\})$-independent multiplicative constant. 
Compatibility of equations (\ref{tauell}) 
is a corollary of the elliptic Schlesinger system. 

\subsection{Schlesinger transformations for elliptic isomonodromic deformations}

The natural generalization of the notion of Schlesinger
transformations on the Riemann sphere to the
elliptic case was given in the paper \cite{KMS99}. Starting from any solution of the
elliptic Schlesinger system (\ref{Se1}) with associated function
$\Psi$ satisfying (\ref{lstor}) and (\ref{perPsi}) we construct a new
solution $\hA_j$, $\hPsi$ with eigenvalues $\tht_j$ which differ
from the $t_j$ by integer or half-integer values. In particular, we
will consider the elliptic analog of the elementary Schlesinger
transformation \Ref{tj} on the Riemann sphere.  The following
construction was inspired by the papers \cite{BiBoIt84}, \cite{DJKM83}.

As an  elliptic analog of the function $F(\l)$ from
(\ref{Fsphere}) we shall choose the following ansatz
\baa
F(\l) &=& \f{f(\l)}{\sqrt{\det f(\l)}} \;,\la{Fell}\\
f(\l) &=& \f{1}{2} +\sum_{\A=1}^3 J_\A\, 
w_\A\!\left(\l-\ft12(\l_k\!+\!\l_l)\right)\,\s_\A \;,\nn
\eaa
where the functions $J_\A(\l_j,\mu)$ depend on $G_k$ and $G_l$ and
will be defined below. The elementary elliptic Schlesinger transformation is described by 
the following
\begin{theorem}\la{dresstor}
\cite{KMS99}
Let the functions $\left\{A_j(\{\l_i\})\right\}$ satisfy the elliptic
Schlesinger system (\ref{Se1}) with twist properties (\ref{Taktwist})
and let the function $\Psi$ satisfy the associated linear system
(\ref{lstor}). For two arbitrary non-coinciding poles $\l_k$ and
$\l_l$, define the new function
\be
\hPsi(\l)~\equiv~ F(\l)\,\Psi(\l) \;,
\la{Psihell}\ee
where $F(\l)$ is given by formula (\ref{Fell}) and $\l$-independent
coefficients $J_\A$ are defined by  
\be
\sum_\A J_\A\, w_\A\!\left(\ft12\,(\l_k\!-\!\l_l)\right)\,\s_\A ~\equiv~  
- \ft{1}{2}\, G \,\s_3\,  G^{-1} \;;
\la{SGell}\ee
as above we denote by $G$ the matrix \Ref{Gkl} containing
the first columns of the matrices $G_k$ and $G_l$. 

Then the function $\hPsi(\l)$  satisfies the equations (\ref{lstor}),
(\ref{Psigjtor}), (\ref{Psitau}) and the twist conditions
(\ref{perPsi}) with the transformed functions 
\be
\hA_j
\left(\{\l_i\}\right)~\equiv~
\res_{\l=\l_j}\left\{\f{d\hPsi}{d\l}\hPsi^{-1}\right\} \;. 
\la{hAjell}\ee
In turn, the functions $\hA_j$ satisfy the elliptic Schlesinger system
(\ref{Se1}). For the eigenvalues $t_j$ we have
\ben
\tht_j~=~ \left\{ \begin{array}{ll} t_j\!+\!\ft12\;\quad &
                               \mbox{for~} j=k,l \\  
                               t_j &\mbox{else} 
                               \end{array} \right. \;.
\een
The monodromy matrices $\hM_j$, $\hM_a$ and $\hM_b$ of
the function $\hPsi$ coincide with the monodromies of $\Psi$, except
for $\hM_k=-M_k$ and  $\hM_l=-M_l$. 
\end{theorem}
\medskip

{\it Proof.} 
The proper local behaviour of function $\hPsi$ at singularities 
$\l_j$ is ensured by the
relations  
\be\la{Sell}
S_\pm^2=S_\pm\;,\qquad S_++S_-=I\;,\qquad
S_+\,G_l^1= S_-\,G_k^1 =0\;;
\ee
for
\ben
S_\pm ~\equiv~ \f{1}{2}\mp \sum_\A J_\A\; 
w_\A\!\left(\ft12\,(\l_k\!-\!\l_l)\right)\, \s_\A ~=~
G \,P_\pm\,  G^{-1} \;,
\een
which in complete analogy to (\ref{eqS}) describe annihilation of the
vectors $G_k^1$ and $G_l^1$ by the matrices $f(\l_k)$ and $f(\l_l)$,
respectively. Obviously, equations \Ref{Sell} are a consequence of
\Ref{SGell}. Similarly to the case of the sphere, it is then easy to
verify that (\ref{Sell}) provide the required asymptotical expansions
(\ref{asymp}) for the function $\hPsi$ with parameters $\hG_j$, $C_j$
and $\tht_j$. 

Concerning the global behavior of $\hPsi$ we note that the prefactor
$(\det f(\l))^{-1/2}$ in (\ref{Fell}) provides the condition
$\det\hPsi =1$ and kills the simple pole of $f(\l)$ at
$\l=(\l_k\!+\!\l_l)/2$. Therefore, the only singularities of $F(\l)$
on $E$ are the zeros of $\det f(\l)$. Since $\det f(\l)$ has only one
pole -- this is the second order pole at $\l=(\l_k\!+\!\l_l)/2$ -- it
must have also two zeros on $E$ whose sum according to Abel's theorem
equals $\l_k+\l_l$.  According to (\ref{Sell}) these are precisely
$\l_k$ and $\l_l$.  It remains to check that $\hPsi$ satisfies
conditions (\ref{perPsi}) with the same matrices $M_a$ and $M_b$.
This follows from the twist properties
\ben
f(\l+1)~=~ \s_3 \,f(\l)\, \s_3\;,\qquad 
f(\l+\mu)~=~ \s_1\, f(\l) \,\s_1\;,
\een
which in turn follow from (\ref{Fell}) and the periodicity
properties (\ref{perw}) of the functions $w_j(\l)$. 

\qed

\medskip

As a result of rather long calculations one can prove the elliptic analog of
formula (\ref{tauh}) describing the transformation of the
$\tau$-function under the action of elliptic Schlesinger
transformations. 

\begin{theorem} \cite{KMS99}
The $\tau$-function $\htau$ corresponding to the Schlesinger-transformed
solution $\hA_j$ (\ref{hAjell}) of the elliptic Schlesinger system is
related to the $\tau$-function corresponding to the solution $A_j$ as
follows 
\be\la{tauhtau}
\htau\left(\{\l_j\},\mu\right)~=~ 
\left\{\Big[w_1 w_2 w_3\!\left(\f{\l_k\!-\!\l_l}{2}\right)\; 
\Big]^{1/2}\;\det\,\Big[ G J^{1/2}\Big]\:\right\}\,
\cdot\, \tau\left(\{\l_j\},\mu\right) \;,
\ee
where $G$ is the matrix \Ref{Gkl} containing the first columns of
the matrices $G_k$, $G_l$,
$$
J\equiv \sum_{\A=1}^3 J_A \sigma_A
$$
 and the functions $J_\A$ are defined in terms of  $G$ via (\ref{SGell}).  
\end{theorem}

The natural open  problem  arising here is to construct elliptic generalizations of integrable chains
associated to ordinary Schlesinger system \cite{JimMiw81b,JimMiw81c}.

In the next section we shall present the extension of construction of algebro-geometric 
solutions of Schlesinger system to the case of elliptic isomonodromic deformations.

\subsection{Algebro-geometric solutions of elliptic Schlesinger system}

To construct theta-functional solutions of elliptic Schlesinger system (\ref{Se1}) 
let us assume that $N=2g$ and introduce two-sheet covering $\L$ of torus $E$ with branch points
$\l_1,\dots\l_{2g}$. Genus of $\L$ equals $g+1$. 
Denote by $*$ the involution of $\L$ interchanging the sheets of the covering.
Let us choose the canonical basis of cycles on $\L$ in such a way 
(see figure 6.2 on p.215 of \cite{Bob?}) that
$$
a_1^* = - a_{g+1}\;, \hskip0.8cm 
b_1^* = - b_{g+1}
$$
$$
a_j^* = - a_j\;, \hskip0.8cm 
b_j^* = - b_j\;, \hskip0.8cm j=2,\dots, g\;.
$$
The basic holomorphic differentials $dU_1,\dots, dU_{g+1}$ on  $\L$ normalized by
$$
\oint_{a_j} dU_k=\delta_{jk}\;,\hskip0.8cm j,k=1,\dots,g+1
$$                   
transform as follows under the action of  involution $*$:
\be
dU_1 (P^*) = - dU_{g+1}(P)\hskip1.0cm
dU_j(P^*) = -dU_j(P)\;,\hskip0.7cm j=2,\dots, g
\la{invU}\ee
Let us introduce the following Prym differentials $dV_j,\;j=1,\dots, g$:
\be
dV_1=\f{1}{2}(dU_1 + dU_{g+1})\hskip0.7cm
dV_j = dU_j\; \hskip0.7cm  j=2,\dots, g
\la{Prym}\ee
and symmetric $g\times g$ matrix of their $b$-periods:
\be
\Pi_{jk} = \oint_{b_j} dV_k
\ee
which has positively-defined imaginary part.
\begin{remark}\rm
Differentials $dV_j$ and matrix $\Pi$ were first introduced by Bobenko \cite{Bob?} in the studies of
classical tops admitting  elliptic Lax representation. These objects are related to standard Prym 
differentials $dW_j$ and standard Prym matrix $\Pi^{Prym}$ as follows: $dW_1= 2dV_1$, $dW_j= dV_j\;, j=2,\dots,g$;
$\Pi= 2 S \Pi^{Prym} S$ where $S$ is diagonal matrix $S = {\rm diag}(\f{1}{2},1,\dots, 1)$.
\end{remark}

 Denote by $\Lh$ the universal covering of curve $\L$.

\begin{theorem}  \la{TEPhi}
Define $2\times 2$ matrix-valued function $\Phi(P)$ on $\Lh$ by the formulas
\be
\Phi(P) = \Phi(P)=\left(\ba{cc}\phi(P)\;\;\;\;\; \phi(P)\\
                  \psi(P)\;\;\;\;\; \psi(P)\ea\right),
\la{EPhi}\ee
where
\be
\phi(P)= \Th \left[^\pb_{\qb}\right]\left(\int_{\l_1}^P dV\Big|\Pi\right) \;,  \hskip0.8cm
\psi(P)= \Th \left[^{\hskip0.3cm \pb}_{\qb+\f{1}{2}\eb_1}\right]\left(\int_{\l_1}^P dV\Big|\Pi\right)\;;
\la{Ephipsi}\ee
$\pb, \qb\in\C$ are arbitrary constant vectors   such that $p_1=0$.
Then the function $\Phi$ is holomorphic and invertible on $\L$ outside of branch points $\l_j$ and transforms
as follows with respect to analytical continuation  along the basic cycles of $\L$:
\be
T_{a_1}[\Phi(P)] = \sigma_1 \Phi(P) 
\la{tr1}\ee
\be
T_{b_1}[\Phi(P)] =\sigma_3 \Phi(P) e^{-2\pi i q_1\sigma_3} e^{-\pi i\Pi_{11}-2\pi i \int_{\l_1}^P dV_1}\\
\la{tr2}\ee
\be
T_{a_j}[\Phi(P)] = \Phi(P) e^{2\pi i p_j\sigma_3}
\la{tr3}\ee
\be
T_{b_j}[\Phi(P)] = \Phi(P) e^{-2\pi i q_j\sigma_3} e^{-\pi i\Pi_{jj}-2\pi i \int_{\l_1}^P dV_j}
\la{tr4}\ee
for $ j=2,\dots,g$.

\end{theorem}

{\it Proof.}
Taking into account the definition of Prym differentials $dV_j$ (\ref{Prym}) we see that
$$
T_{a_1}\left[\int_{\l_1}^P dV\right]= \int_{\l_1}^P dV +\f{\eb_1}{2}
$$
$$
T_{a_j}\left[\int_{\l_1}^P dV\right]= \int_{\l_1}^P dV +\eb_1\;, \hskip0.8cm j=2,\dots, g
$$
$$
T_{b_j}\left[\int_{\l_1}^P dV\right]= \int_{\l_1}^P dV + \Pi\eb_j\;, \hskip0.8cm j=1,\dots, g
$$
Substituting these expressions into the formulas for $\phi$ and $\psi$ and taking into account 
behaviour of 1-forms $dU_j$ under the action of involution $*$ (\ref{invU}) we derive the following
transformation properties of functions $\phi$ and $\psi$:
\be
T_{a_1}[\phi(P)] = \psi(P)\hskip0.8cm
T_{a_1}[\psi(P)] = \phi(P)
\ee
\be
T_{a_1}[\phi(P^*)] = \psi(P^*)\hskip0.8cm
T_{a_1}[\psi(P^*)] = \phi(P^*)
\ee
\be
T_{b_1}[\phi(P)] = e^{-2\pi i q_1} e^{-\pi i \Pi_{11}-2\pi i \int_{\l_1}^P dV_1}\phi(P)
\ee
\be
T_{b_1}[\psi(P)] = -e^{-2\pi i q_1} e^{-\pi i \Pi_{11}-2\pi i \int_{\l_1}^P dV_1}\psi(P)
\ee
\be
T_{b_1}[\phi(P^*)] = e^{2\pi i q_1} e^{-\pi i \Pi_{11}-2\pi i \int_{\l_1}^P dV_1}\phi(P^*)
\ee
\be
T_{b_1}[\psi(P^*)] = -e^{2\pi i q_1} e^{-\pi i \Pi_{11}-2\pi i \int_{\l_1}^P dV_1}\psi(P^*)
\ee
and 
\be
T_{a_j}[\phi(P)] = e^{2\pi i p_j} \phi(P)
\ee
\be
T_{a_j}[\phi(P^*)] = e^{-2\pi i p_j} \phi(P^*)
\ee
\be
T_{b_j}[\phi(P)] =  e^{-2\pi i q_j} e^{-\pi i \Pi_{jj}-2\pi i \int_{\l_1}^P dV_j} \phi(P)
\ee
\be
T_{b_j}[\phi(P^*)] =  e^{2\pi i q_j} e^{-\pi i \Pi_{jj}-2\pi i \int_{\l_1}^P dV_j} \phi(P^*)
\ee
Transformation laws of $\psi$ along cycles $a_j, b_j$ for $j>1$  coincide with transformation laws
of $\phi$.

Combining the above relations into matrix form, we come to the transformation laws (\ref{tr1}) - (\ref{tr4}).

It remains to verify non-degeneracy of $\Phi(P)$ outside of singularities $\l_j$. We know 
that  $\det\Phi(P)$ has at least  simple zeros at the points $\l_j$ (at these points the columns of $\Phi(P)$
are proportional to each other). 
To check that  $\det\Phi(P)$ does not vanish on $\Lh$ outside of $\l_j$ let us 
first observe that 
\be
T_{a_1}[\det\Phi(P)] = -\det \Phi(P) \hskip0.8cm 
T_{a_j}[\det\Phi(P)] = \det \Phi(P)\;,\hskip0.6cm j=2,\dots,g\;,
\la{trdet1}\ee
\be
T_{b_j}[\det\Phi(P)] = e^{-2\pi i \Pi_{jj}-4\pi i \int_{\l_1}^P dV_j}\det\Phi(P)\hskip0.6cm j=1,\dots,g\;.
\la{trdet2}\ee

Now let us  calculate the integral
\be
\oint_{\p\Lh} d\log\det\Phi(P) 
\ee
$$
=\oint_{a_1} d\left\{\log\det\Phi(P) -\log\det T_{b_1}[\Phi(P)]\right\} +
\oint_{a_{n+1}} d\left\{\log\det\Phi(P) -\log\det T_{b_{n+1}}[\Phi(P)]\right\} 
$$
$$
+\sum_{j=2}^g \oint_{a_j} d\left\{\log\det\Phi(P) -\log\det T_{b_1}[\Phi(P)]\right\}\;.
$$
Taking into account (\ref{trdet1}) and  (\ref{trdet2}) as well as normalization of the 
basic integrals $dU_j$ we see that the first two terms of the r.h.s. of this expression equal $2\pi i$,
whereas each term in the sum equals $4\pi i$. Altogether, we get $4\pi i g$, and, therefore, $\det\Phi(P)$
has in $\Lh$ exactly $2g$ zeros which coincide with $\l_j$.
\vskip0.8cm

Let us also choose some domain $\Omega\subset E$ which does not overlap with projections of all basic
cycles on $E$. Then domain  $\Omega^*$ does not overlap with the boundary of $\Lh$ and functions 
$\phi^*(P)$ and $\psi^*(P)$ are uniquely defined in $\Lh$ by (\ref{Ephipsi}). 
Let us now choose some sheet of universal covering $X$ of torus $E$ with punctures $\{\l_1,\dots,\l_{2g}\}$,
and define new function $\Psi(\l)$ in subset $\Omega$ of this sheet by the formula
\be
\Psi(\l\in\O)=\f{1}{\sqrt{\det\Phi(\l)}}\Phi(\l)\;.
\la{EPsi}\ee
Then we extend function $\Psi(\l)$ on the rest of $X$ by analytical continuation.

The following theorem shows that function $\Psi$ satisfies conditions of isomonodromy, and, therefore,
generates a class of solutions of elliptic Schlesinger system (\ref{Se1}):
\begin{theorem}
Function $\Psi(\l\in X)$ defined by formulas (\ref{EPhi}), (\ref{Ephipsi}) and  (\ref{EPsi}) is holomorphic and
invertible on $X$ outside of the points $\l_j,\; j=1,\dots, 2g$. Moreover, it transforms
as follows with respect to analytical continuation along basic cycles of $E$:
\be
T_a[\Psi(\l)] = i\sigma_1 \Psi(\l)\hskip0.8cm
T_b[\Psi(\l)] = i\sigma_3 \Psi(\l) e^{-2\pi i q_1\sigma_3}
\la{Psiab}\ee
and around closed cycles surronding   points $\l_j$:
\be
T_{\l_j}[\Psi(\l)] = \Psi(\l) M_j
\la{Psigj}\ee
where $T_b$, $T_b$ and $T_{\l_j}$ denote corresponding operators of analytical continuation;
\be
M_j= \left(\ba{cc} 0 & -m_j \\m_j^{-1} & 0 \ea\right)\;,
\la{EMj}\ee
and 
\be
m_1 = i\hskip1.0cm 
m_2 = -i\exp\{-2\pi i\sum_{j=2}^g p_j\}
\la{Em1m2}\ee
\be
m_{2l}=  -i\exp\{2\pi i q_l -2\pi i \sum_{k=l+1}^g p_k\}
\la{Emeven}\ee
\be
m_{2l-1}=  i\exp\{2\pi i q_l - 2\pi i\sum_{k=l}^g p_k\}
\la{Emodd}\ee
for $l=2,\dots, g$.
\end{theorem}

{\it Proof.} Holomorphy and invertibility of function $\Psi$ follows from the same statements concerning function
$\Phi$ (\ref{EPhi}). Relations (\ref{Psiab}) directly follow from (\ref{tr1}), (\ref{tr2}).
 To calculate $m_j$ let us observe that, according to (\ref{tr1}) -  (\ref{tr4}),
monodromies of $\Psi$ are related to constants $\pb$ and $\qb$ as follows:
\be
M_{2j} M_{2j-1} = e^{2\pi i p_j\sigma_3}\;,\hskip0.8cm j=2,\dots, g
\ee
\be
M_{2j-1} M_{2j-2} = e^{2\pi i (q_j-q_{j-1})\sigma_3}\;,\hskip0.8cm j=3,\dots, g
\ee
\be
M_3 M_2 = e^{2\pi i q_2}
\ee
Moreover, we order monodromies is such a way that
\be
M_a M_b M_a^{-1} M_b^{-1} M_{2g}\dots M_1 =I
\ee
(since $M_a=I$ first four factors in this relation drop out). 
Monodromy $M_1$ corresponding to our choice of basic cycles 
equals $i\sigma_1$. 
Altogether all these relations lead to (\ref{EMj}),  (\ref{Em1m2}) - (\ref{Emodd}) after elementary
calculations. \qed

\begin{corollary}
Residues
\be
A_j(\{\l_j\},\mu)\equiv {\rm res}|_{\l=\l_j} \Psi_\l\Psi^{-1}
\ee
satisfy elliptic Schlesinger system (\ref{Se1}).
\end{corollary}

\section{Outlook}

Let us mention several applications of the mathematical results  described above.
Recently \cite{KorNic95} 
it was established close relationship between Schlesinger system and Ernst equation of
general relativity, which allows to apply to the Ernst equation all results of sect.2. 
In particular, one can get in this way a class of algebrogeometric solutions of Ernst equation
\cite{KorMat98}, which turns out to coincide with the class of algebrogeometric solutions
of Ernst eqation known since 1988 \cite{Koro88}. It is rather satisfactory that
certain subclass of genus 2
 algebrogeometric solutions of Ernst equation recently found realistic physical application in the 
problem of description of different kinds of dust discs \cite{NeuMei94,KleRic99}.
Another application of construction of sect.2 is the theory of $SU(2)$-invariant gravitational
instantons \cite{BabKor98} where it allows to considerably simplify the results of  Hitchin
\cite{Hitc94}.

So far we don't know about physical applications of elliptic version of Schlesinger system,
and all results of sect.3 have at the moment pure mathematical significance; however we strongly believe
that such applications will be found in the near future.

\begin{appendix}
\section{Some elliptic functions}
 
The elliptic theta-function with characteristic $[p,q]$ ($p,q\in \C$)
on a torus $E$ is defined by the series
\be
\th[p,q](\l|\mu)=\sum_{m\in \Z} e^{\pi \i\mu (m+p)^2 +2\pi \i 
(m+p)(\l+q)} \;. 
\ee
Let us introduce on the torus $E$ the standard Jacobi theta-functions:
\baa
\th_1(\l)&\equiv& - \th\left[\ft{1}{2},\ft{1}{2}\right](\lambda|\mu) \;,
\\
\th_2(\l)&\equiv& \th\left[\ft{1}{2},0\right](\lambda|\mu) \;,
\non
\th_3(\l)&\equiv&\th(\l)\equiv  \th[0,0](\lambda|\mu) \;,
\non
\th_4(\l)&\equiv& \th\left[0,\ft{1}{2}\right](\lambda|\mu)\;,
\nn
\eaa
and corresponding theta-constants
\ben
\th_j\equiv \th_j(0)\;,\hskip0.8cm j=2,3,4\;.
\een
We define the following three combinations of Jacobi theta-functions:
\be
w_1(\l)= \pi\th_2\th_3 \f{\th_4(\l)}{\th_1(\l)}\;,\quad
w_2(\l)= \pi\th_2\th_4 \f{\th_3(\l)}{\th_1(\l)}\;,\quad
w_3(\l)= \pi\th_3\th_4 \f{\th_2(\l)}{\th_1(\l)}\;.
\la{wj}\ee
All these functions have simple poles at $\l=0$ with residue
1. Moreover, they possess the following periodicity properties: 
\baa
w_1(\l+1) = - w_1(\l) \quad&&\qquad w_1(\l+\mu) = \phantom{-}w_1(\l)\;, 
\la{perw} \\
w_2(\l+1) = - w_2(\l) \quad&&\qquad w_2(\l+\mu) = - w_2(\l)\;, \non
w_3(\l+1) = \phantom{-}  w_3(\l) \quad&&\qquad w_3(\l+\mu) = - w_3(\l)\;.
\nn
\eaa
Let us also define the following functions $\Zf_\A$:
\be
\Zf_1= \f{w_1}{2\pi i}\f{\th_4'(\l)}{\th_4(\l)} \;,\quad
\Zf_2= \f{w_2}{2\pi i}\f{\th_3'(\l)}{\th_3(\l)}\;,\quad
\Zf_3= \f{w_3}{2\pi i}\f{\th_2'(\l)}{\th_2(\l)}\;.
\la{Z}\ee
which have the  periodicity properties:
\baa
\Zf_1 (\l+1) = - \Zf_1 (\l) \quad&&\qquad 
\Zf_1 (\l+\mu) = \phantom{-}\Zf_1 (\l) - w_1\;, \la{perZ} \\ 
\Zf_2 (\l+1) =  - \Zf_2 (\l) \quad&&\qquad 
\Zf_2 (\l+\mu) =  - \Zf_2 (\l) + w_2\;, \non
\Zf_3 (\l+1) = \phantom{-} \Zf_3 (\l) \quad&&\qquad 
\Zf_3 (\l+\mu) =  - \Zf_3 (\l) + w_3\;. \nn
\eaa
It is easy to verify the identity
\be
\f{d w_\A}{d\mu}(\l)= \f{d \Zf_\A}{d\l}(\l)\;,
\la{cross}\ee
which follows from analyticity and twist properties of both sides.

Notice also that functions $w_\A$ may be represented as ratios of Jacobi's elliptic
functions as follows: 
\be
w_1(\l)= \f{1}{{\rm sn}(\l)\;}\;,\quad 
w_2(\l)= \f{{\rm dn}(\l)}{{\rm sn}(\l)}\;,\quad 
w_3(\l)= \f{{\rm cn}(\l)}{{\rm sn}(\l)} \;.
\la{wsn}\ee

In calculation of transformation of elliptic $\tau$-function under the action of elliptic 
Schlesinger transformations one has to use also the summation theorem and some 
integral relations for functions $w_\A$ \cite{KMS99}.

\end{appendix}


\begin{thebibliography}{10}

\bibitem{Schl12}
L.~Schlesinger, 
{\em J. Reine u.
  Angew. Math.} {\bf 141} (1912) 96--145.

\bibitem{JiMiMoSa80}
M.~Jimbo, T.~Miwa, Y.~M\^{o}ri, and M.~Sato, 
{\em Physica} {\bf 1D} (1980) 80--158.

\bibitem{JiMiUe81}
M.~Jimbo, T.~Miwa, and K.~Ueno, 
{\em Physica} {\bf 2D} (1981) 306--352.

\bibitem{JimMiw81b}
M.~Jimbo and T.~Miwa, 
{\em Physica} {\bf
  2D} (1981) 407--448.

\bibitem{JimMiw81c}
M.~Jimbo and T.~Miwa, 
{\em Physica} {\bf
  2D} (1981) 26--46.

\bibitem{KitKor98}
A.~Kitaev and D.~Korotkin, 
{\em Intern. Math. Research Notices} {\bf 17}
  (1998) 877--905

\bibitem{DIKZ99}
P.Deift, A.Its, A.Kapaev, and X.Zhou, 
{\em Commun. Math. Phys.} 
{\bf 203} (1999) 613--634.

\bibitem{Okam71} 
K.~Okamoto, 
{\em
Funkcial. Ekvac.}  {\bf 14} (1971) 137--152; 
{\em J. Fac.  Sci. Univ. Tokyo Sect. IA
Math.} {\bf 24} (1977) 357--372.

\bibitem{Okam79}
K.~Okamoto, 
{\em J. Fac. Sci. Univ.
  Tokyo Sect. IA Math.} {\bf 26} (1979) 501--518.

\bibitem{Iwas91}
K.~Iwasaki, 
{\em J. Fac. Sci. Univ. Tokyo
Sect. IA Math.} {\bf 38} (1991) 431--531; 
{\em Pacific J. Math.} {\bf 155}
(1992) 319--340

\bibitem{KorSam97}
D.~Korotkin and H.~Samtleben, 
{\em Int. J. Mod. Phys.} {\bf A12} (1997)
  2013--2030 

\bibitem{Taka98}
K.~Takasaki, 
{\em Lett. Math. Phys.} {\bf 44} (1998) 143--156
 

\bibitem{KMS99} D.Korotkin, N.Manojlovi\'c, H.Samtleben, {\it Schlesinger 
transformations for elliptic isomonodromic deformations},
solv-int/9910010


\bibitem{HarIts97}
J.Harnad and A.Its, {\it Integrable {F}redholm operators and dual isomonodromic
  deformations, {P}reprint {CRM}-2477, {M}ontr\'eal},
 solv-int/9706002

\bibitem{Kita97}
A.~Kitaev, {\it Special functions of the isomonodromy type},  {\em Preprint
  SFB-288-272, Berlin} (1997, 27pp.).

\bibitem{Fay} John D. Fay, Theta functions on Riemann surfaces, Lecture Notes in Mathematics,
{\bf 352}, Springer, 1973



\bibitem{Etin94}
P.~I. Etingof, 
{\em Commun. Math. Phys.} {\bf
  159} (1994) 471--502


\bibitem{BiBoIt84}
R.Bikbaev, A.Bobenko, and A.Its, {\it Theory of exact solutions of
  {L}andau-{L}ifshitz equation},  {\em Preprint DonFTI 84-6(81) I,II} (1984).

\bibitem{DJKM83}
E.~Date, M.~Jimbo, M.~Kashiwara, and T.~Miwa, 
  {\em J. Phys.} {\bf A16} (1983) 221--236.

\bibitem{Ivan96}
D.Ivanov, {\it Knizhnik-{Z}amolodchikov-{B}ernard equations as a quantization
  of nonstationary {H}itchin system}, Preprint hep-th/9610207

\bibitem{LevOls97}
A.Levin and M.Olshanetskii, {\it Hierarchies of isomonodromic
  deformations and {H}itchin systems}, Preprint {ITEP-TH} 45/97,
  hep-th/9709207

\bibitem{Bob?} E.Belokolos, A.Bobenko, V.Enolskii, A.Its, V.Matveev, 
{\it Algebro-geometric approach to nonlinear integrable equations},
Springer series in non-linear dynamics, Berlin Heidelberg New York, 1994


\bibitem{KorNic95} D.Korotkin, H.Nicolai,{\em  Phys.Rev.Lett.},  74 1272-1275 (1995)


\bibitem{KorMat98} D.Korotkin, V.Matveev, {\it On theta-functional solutions of Schlesinger system 
and Ernst equation}, gr-qc/9810041


\bibitem{Koro88}  D.Korotkin, {\em Theor.Math.Phys.}  {\bf 77} 1018-1031 (1989)


\bibitem{NeuMei94}G.Neugebauer and R.Meinel,  {\em Phys.Rev.Lett.} {\bf 75}  3046-3048 (1995)


\bibitem{KleRic99} C.Klein and O.Richter, {\em Phys.Rev.D} {\bf 57} 857-862 (1998)

\bibitem{BabKor98} M.Babich and D.Korotkin,  {\em Lett.Math.Phys.} {\bf 46}: 323-337 (1998)


\bibitem{Hitc94}N.Hitchin, {\em J.Diff.Geom.}, {\bf 42} (1) 30-112 (1995)



\end{thebibliography}
\end{document}